\documentclass[prd,nofootinbib,showpacs]{revtex4}
\usepackage{graphicx}
\usepackage[usenames,dvipsnames]{color}
\usepackage{amsmath,amssymb}
\usepackage{epstopdf}
\usepackage{hyperref}

\begin{document}

\title{Spontaneous CP violating quark scattering 
from asymmetric $Z\left(3\right)$ interfaces in QGP}

\author{Abhishek Atreya}
\email{atreya@iopb.res.in}
\author{Partha Bagchi}
\email{partha@iopb.res.in}
\author{Arpan Das}
\email{arpan@iopb.res.in}
\author{Ajit M. Srivastava}
\email{ajit@iopb.res.in}
\affiliation{Institute of Physics, Bhubaneswar, 751005, India}

\begin{abstract}
In this paper, we extend our earlier study of spontaneous CP violating
scattering of quarks and anti-quarks from QCD $Z\left(3\right)$ domain 
walls for the situation when these walls have asymmetric profiles of
the Polyakov loop order parameter $l(x)$. Dynamical quarks lead to 
explicit breaking of $Z(3)$ symmetry, which lifts the degeneracy of the 
$Z(3)$ vacua arising from spontaneous breaking of the $Z(3)$ symmetry in 
the quark-gluon plasma (QGP) phase. Resulting domain walls have 
asymmetric profile of $l(x)$ (under reflection $x \rightarrow -x$ for a 
domain wall centered at the origin). We calculate the background gauge 
field profile $A_0$ associated with this domain wall profile. Interestingly,
even with the asymmetric $l(x)$ profile, quark-antiquark scattering from 
the corresponding gauge field configuration does not reflect this asymmetry.
We show that the expected asymmetry in scattering arises when we include 
the effect of asymmetric profile of $l(x)$ on the effective mass of 
quarks and antiquarks and calculate resultant scattering. We discuss the 
effects of such asymmetric Z(3) walls in generating quark and antiquark 
density fluctuations in cosmology, and in relativistic heavy-ion collisions 
e.g. event-by-event baryon fluctuations. 
\end{abstract}

\maketitle

\section{Introduction}
\label{sec:intro} 

 Search for the quark-gluon plasma (QGP) phase of QCD in relativistic
heavy-ion collision experiments (RHICE) has reached a mature stage
with observations providing compelling evidence that QGP phase is
created in these experiments. While a definitive conclusion about the
discovery of QGP is still awaited, it is an appropriate stage to 
explore new effects and new structures in this exotic phase of QCD.
This is particularly important due to its implications for the case
of early universe as well as for the cores of dense astrophysical
objects like neutron stars. One such new effect is the possibility of 
extended topological objects in the QGP phase which arise from 
spontaneous breaking of the center symmetry $Z(3)$ of the color 
$SU(3)$ group. The $Z(3)$ symmetry is broken spontaneously as the Polyakov 
loop, $l(x)$, which is an order parameter for the  
confinement-deconfinement phase transition for pure gauge theory, 
\cite{Polyakov:1978vu} assumes a non-zero value in the deconfined phase. 
The resulting domain walls, so called $Z(3)$ walls 
\cite{Bhattacharya:1992qb,West:1996ej,Boorstein:1994rc}, are in some 
sense similar to the axionic domain walls in the universe. Interestingly,
just like axionic cosmic strings, here also there are topological
strings associated with the junctions of these $Z(3$) walls \cite{Layek:2005fn}.
The study of these defects becomes more relevant in the present 
era of relativistic heavy-ion collision experiments as the temperature and 
energy densities that are needed to form these defects is (hopefully) 
accessible in these accelerators. In fact, these defects are the only 
defects in a relativistic quantum field theory that can be probed in the 
present day laboratory conditions.

In earlier works, some of us have studied the formation and evolution of 
these topological objects in the initial transition to the QGP phase in 
the context of RHICE \cite{Gupta:2010pp}. Various consequences of
Z(3) walls have been discussed in these works for RHICE arising from
nontrivial scattering of quarks from Z(3) walls. Implications of
the existence of these walls in the early universe has also been discussed in 
\cite{Layek:2005zu} where it is shown that baryon inhomogeneities can
arise from scattering of quarks from Z(3) walls. Scattering of 
quarks/antiquarks was studied  in \cite{Layek:2005zu} by modeling the 
dependence of effective quark mass on the magnitude of the Polyakov loop
order parameter $l(x)$. Spatially varying profile of $l(x)$ leads to
spatially varying effective mass, which behaves as potential
in the Dirac equation for quarks/antiquarks leading to non-trivial 
scattering. As this effective mass (potential) is the same for quarks and
antiquarks, resulting scattering is the same for both.

  In \cite{Atreya:2011wn} we followed a different method for studying the 
scattering
of quarks/antiquarks from Z(3) walls. We assume that the profile of $l(x)$
corresponds to a sort of condensate of the background gauge field $A_0$
(following the definition of the Polyakov loop order parameter). We
calculate this profile of the background gauge field from the profile of 
$l(x)$. Such a gauge field configuration, when used in the Dirac equation,
leads to a potential which is different for quark and antiquark, leading
to spontaneous CP violation in the scattering of quarks and antiquarks from
a given Z(3) wall. This CP violation is spontaneous as it arises from a
specific background configuration of the gauge field corresponding to 
a given Z(3) wall. This was first discussed by Altes et al. 
\cite{KorthalsAltes:1994be,KorthalsAltes:1994if} who argued in the context of 
the universe, that 
due to the non-trivial background field configuration for the standard 
model gauge fields, the localization of quarks and antiquarks on the wall 
is different. Its possible effects on the electroweak baryogenesis via 
sphalerons was discussed in \cite{KorthalsAltes:1994be,KorthalsAltes:1994if}. 
This spontaneous CP 
violation for the case of QCD was also discussed in 
\cite{KorthalsAltes:1992us}. The CP violating 
effects discussed in above works were primarily qualitative, as 
the exact profiles of $A_0$ were not calculated. 
In \cite{Atreya:2011wn}, we use the profile of 
Polyakov loop $l(x)$ between different Z(3) vacua (which was obtained by 
using specific effective potential for $l(x)$ as discussed in 
\cite{Pisarski:2000eq}) to obtain the full profile of the background gauge 
field $A_0$. This background $A_0$ configuration acts as a potential for 
quarks and antiquarks causing non-trivial reflection of quarks from the wall. 
There we also showed that this spontaneous CP violation arising 
from the background $A_0$ configuration leads to different reflection 
coefficients for quarks and antiquarks. In a series of follow up works 
\cite{Atreya:2014sea,Atreya:2014sca}, we studied 
the effect of this difference in the scattering of quarks and antiquarks 
from $Z(3)$ walls in the context of ongoing relativistic heavy-ion collision
experiments and the early universe. In \cite{Atreya:2014sea}, we discussed 
a novel mechanism of $J/\psi$ disintegration in the relativistic 
heavy ion collision experiments. We showed that the localized electric field
in the CP violating Z(3) domain wall in the QGP phase lead to disintegration 
of quarkonia. In \cite{Atreya:2014sca}, we studied the effect of this CP 
violation on baryon transport across the collapsing $Z(3)$ domain 
walls in the early universe. We showed that it can 
lead to formation of quark nuggets as well as antiquark nuggets by segregating 
baryons and antibaryons in different regions of the universe near QCD phase 
transition epoch. As quarks are concentrated in a given collapsing domain 
wall, similar amount of antiquarks get concentrated in another collapsing 
domain wall which has the CP conjugate configuration of $A_{0}$ corresponding 
to the interchange of the two $Z(3)$ vacua with respect to the first domain 
wall case. Thus, for a given size of collapsing domain walls, resulting
nugget sizes are identical for quarks and antiquarks.

There have been some objections on the existence of Z(3) walls, (and
of the associated field $A_{0}$) in the 
Minkowski space. We refer to our earlier work \cite{Atreya:2011wn} for 
a discussion on this aspect. In our above mentioned 
studies of CP violating scattering of quarks/antiquarks from $Z(3)$ walls 
we have neglected the effect of quarks. The existence of these $Z(3)$ walls 
becomes a non-trivial issue in the presence of quarks. It has been argued 
that $Z(3)$ symmetry loses it's meaning in the presence of dynamical quarks
\cite{Smilga:1993vb,Belyaev:1991np}. Another approach to this issue is to 
regard the effect of quarks in terms of the explicit breaking of Z(3) symmetry 
\cite{Dumitru:2000in,Dumitru:2001bf,Dumitru:2002cf}. This finds support in the 
recent lattice calculations of QCD with quarks \cite{Deka:2010bc}, which 
suggest that there is a strong possibility of existence of these $Z(3)$ vacua
 at high temperature. Since the presence of quarks lifts the
degeneracy of different $Z(3)$ vacua, the
$Z(3)$ interfaces are no more solutions of time independent field equations
as they move away from the region with the unique true vacuum. 
However, it is important to note that with quark effects (taken in terms 
of explicit symmetry breaking), the interfaces survive as non-trivial 
topological structures, even though they do not remain solutions of 
time independent equations of motion. As the resulting profile of $l(x)$ 
between the true vacuum and a metastable vacuum is no more symmetric, it
raises interesting possibilities for the generation of quark and
antiquark inhomogeneities as a network of collapsing domain walls
is considered, with different walls interpolating between different sets 
of $Z(3)$ vacua. Situation is even more interesting as with explicit
symmetry breaking certain closed domain walls with true vacuum inside 
(and with sufficiently larger size) may expand \cite{Gupta:2011ag}.
This can lead to concentration of quarks and antiquarks in a shell like 
structure, which can have important implications in cosmology (for
large shells) and in RHICE where it may imply concentration of 
baryons or antibaryons near the surface of the QGP region. With these
motivations, we extend our earlier study of \cite{Atreya:2011wn} in this paper
with incorporation of the effects of explicit symmetry breaking arising
from dynamical quarks. We find that even though the profile of
$l(x)$ is asymmetric in this case (under reflection $x \rightarrow -x$)
quark-antiquark scattering from the gauge field configuration associated
with it does not show any difference from the symmetric case when  
explicit $Z(3)$ symmetry breaking is absent. More precisely, the scattering
of a quark from left on the wall is identical to the scattering of an
antiquark from the right. We then include the effect of asymmetric profile 
of $l(x)$ on the effective mass of quarks and antiquarks and calculate 
resultant scattering. Due to asymmetric profile of $l(x)$
the resulting effective mass of quarks and antiquarks is different
when considering scattering from the left or from the right. (Though it
is the same for quark and antiquark.) This, combined with the CP violating
scattering resulting from the background gauge field configuration
associated with this $l(x)$, leads to left-right asymmetry in scattering
of quarks (from left) and antiquarks (from right). This will lead to
important differences in resulting concentrations of quarks and
antiquarks in cosmology as well as in RHICE.

  The paper is organized in the following manner. In section \ref{sec:cpviol}
we recall the basic physics of the origin of spontaneous CP 
violation due to the presence of $Z(3)$ interfaces and briefly introduce the 
effective potential for 
the Polyakov loop incorporating explicit breaking of $Z(3)$ symmetry.
Calculation of the {\it asymmetric} profile of $l(x)$ for this case
and its associated gauge field configuration is somewhat non-trivial 
and we discuss this in section \ref{sec:asymprfl}. In section 
\ref{sec:ref-trans} we first discuss the 
scattering of quarks and antiquarks from this gauge field configuration
and show that it leads to the same results for quark and antiquark
concentrations as for the case without any explicit symmetry breaking. 
We then introduce $l(x)$ dependent effective mass for quarks and antiquarks
and show that the resultant scattering is different for quarks (from
the left) and antiquarks (from the right). Section \ref{sec:discussion} 
presents discussion and conclusions where we discuss possible implications 
of these results for cosmology and for RHICE. 

\section{SPONTANEOUS CP VIOLATION FROM Z(3) WALLS} \label{sec:cpviol}

 We briefly recall the basic physics of the origin of the spontaneous
CP violation arising from $Z(3)$ walls. The source of this CP violation
is a background condensate of the gauge field $A_0$ which we take to 
correspond to the profile of $l(x)$. This association is made following
the definition of the Polyakov loop,
\cite{Polyakov:1978vu,Gross:1980br,McLerran:1981pb}

\begin{equation} 
L(x) = \frac{1}{N}Tr\biggl[\mathbf{P} \exp\biggl(ig\int_{0}^{\beta}A_{0}
  (\vec{x},\tau)d\tau\biggr)\biggr],
\label{eq:lx}
\end{equation}

where $A_{0}(\vec{x},\tau) = A_{0}^{a}(\vec{x},\tau)T^{a}, (a = 1,\dotsc N)$
are the gauge fields and $T^{a}$ are the generators of $SU\left(N\right)$
in the fundamental representation. $\mathbf{P}$ denotes the path ordering in
the Euclidean time $\tau$, and $g$ is the gauge coupling. Under global 
$Z(N)$ symmetry transformation, the Polyakov Loop transforms as
\begin{equation}
L(x) \longrightarrow Z\times L(x), \qquad \textrm{where } Z = e^{i\phi},
\label{eq:zn}
\end{equation}
with $\phi = 2\pi m/N$; $m = 0,1 \dotsc (N-1)$. 

Thermal average of the Polyakov loop, $\langle L(x)\rangle$, (which we
denote as $l(x)$) is related to the free energy of an infinitely heavy 
test quark in a pure gluonic medium ($l(x) \propto e^{-\beta F}$).
In the confined phase, a test quark should have infinite energy implying
that $l(x)$ = 0. In the deconfined phase, a test quark will have finite
energy implying non-zero value of $l(x)$. Thus $l(x)$ serves as an order
parameter for the confinement-deconfinement transition. In view of
Eq.(\ref{eq:zn}), a non-zero value of $l(x)$ leads to spontaneous breaking 
of $Z(N)$ symmetry in the high temperature deconfined phase, while this 
symmetry is restored in the low temperature confined phase when $l(x) = 0$. 
For QCD, $N = 3$, hence confinement-deconfinemnet transition in QCD corresponds
to spontaneous breaking of $Z(3)$ symmetry leading to $Z(3)$ domain walls
(and associated QGP string, see ref.\cite{Layek:2005fn}).

We emphasize that, though certainly there are conceptual issues 
regarding the existence of these structures 
\cite{Smilga:1993vb,Belyaev:1991np}, by no means their
existence can be ruled out. In fact, amongst all models allowing
for existence of topological extended structures (domain walls,
strings etc. which have been proposed in various particle physics
theories in the early universe), these Z(3) domain walls (and the
associated QGP string) are the most well motivated. Indeed, if
these objects exist, these will be the only relativistic field
theory topological solitons which are accessible in  laboratory
experiments. Their detection will not only provide deep insights
in the non-trivial physics of the QGP phase, it will have
very important implications for cosmology. 
 
 As we mentioned, we determined the background gauge field configuration
$A_0$ from the profile of $l(x)$ for a specific domain wall which
interpolates between two $Z(3)$ vacua without quark effects. 
For determining the profile of $l(x)$ interpolating between different 
$Z(3)$ vacua we used the specific effective potential for the Polyakov
loop from ref. \cite{Pisarski:2000eq} with the Lagrangian density given by
\begin{equation}
L=\frac{N}{g^2} \lvert\partial_\mu l\rvert ^2{T^2}- V(l) .
\label{eq:lagden}
\end{equation}
Here $N=3$ for QCD. $T^{2}$ is multiplied with the first term to
give the correct dimensions to the kinetic term. The effective potential 
$V(l)$ for the Polyakov loop is given as
\begin{equation}
V(l) = \biggl(-\frac{b_2}{2}|l|^{2} - \frac{b_3}{6}\Bigl(l^{3} 
+ (l^{*})^{3}\Bigr) + \frac{1}{4}(|l|^{2})^{2}\biggr)b_4T^{4}.
\label{eq:pispot}
\end{equation}
The coefficients $b_2$, $b_3$ and $b_4$ are dimensionless
quantities. These parameters are 
fitted in ref.\cite{Dumitru:2000in,Dumitru:2001bf,Dumitru:2002cf} such that
that the effective potential reproduces the thermodynamics of pure
$SU(3)$ gauge theory on lattice \cite{Boyd:1996bx,Okamoto:1999hi}. The 
coefficients are $b_2 = \left( 1-1.11/x \right) \left(1+0.265/x\right)^{2} 
\left(1+0.300/x\right)^{3} - 0.478$, (with $x=T/T_{c}$ and $T_{c}\sim182$ MeV), 
$b_3 = 2.0$ and $b_4 = 0.6061\times47.5/16$.  With these values, 
$l\left(x\right)\longrightarrow y = b_{3}/2+\frac{1}{2}\times \sqrt{b_{3}^{2} 
+ 4b_{2}\left(T=\infty\right)}$ as $T \longrightarrow \infty$. Various 
quantities are then rescaled such that $l\left(x\right) \longrightarrow 1$ as 
$T \longrightarrow \infty$. The scaling are
\begin{equation}
l\left(x\right) \rightarrow \frac{l\left(x\right)}{y}, ~~ b_{2} \rightarrow 
\frac{b_{2}}{y^{2}}, ~~ b_{3} \rightarrow \frac{b_{3}}{y}, ~~ b_{4} \rightarrow 
b_{4}y^{4}.
\end{equation}
  At low temperature where $l = 0$, the potential has only one
minimum. For temperatures higher than $T_c$, the Polyakov loop
develops a non vanishing  vacuum expectation value $l_0$, and
the cubic term above leads to $Z(3)$ degenerate vacua. The $l\left(x\right)$ 
profile is calculated by energy minimization, see ref.\cite{Layek:2005fn} for 
details. From the $l(x)$ profile, the $A_{0}$ profile is calculated by 
inverting Eq. (\ref{eq:lx}). For various conceptual issues regarding this 
calculation we 
refer to our earlier work \cite{Atreya:2011wn}. To address the issue of 
uncertainties in the determination of the $A_0$ profile depending on the 
choice of the specific form of the effective potential, we had repeated this 
calculation of $A_0$ profile, in ref.\cite{Atreya:2011wn}, for another choice 
of effective potential of the Polyakov loop as provided by 
Fukushima \cite{Fukushima:2003fw}. It was found that even though the two 
effective potentials (in refs.\cite{Pisarski:2000eq} and  
\cite{Fukushima:2003fw}) are of qualitatively different shapes, the 
resulting wall profile and $A_0$ profile were very similar. This gives us 
confidence that our conclusions arising from the calculations of 
scattering of quarks and antiquarks from $Z(3)$ walls are not crucially
dependent on the specific choice of the effective potential.
\begin{figure}[!htp]
\begin{center}
\includegraphics[width=0.45\textwidth]{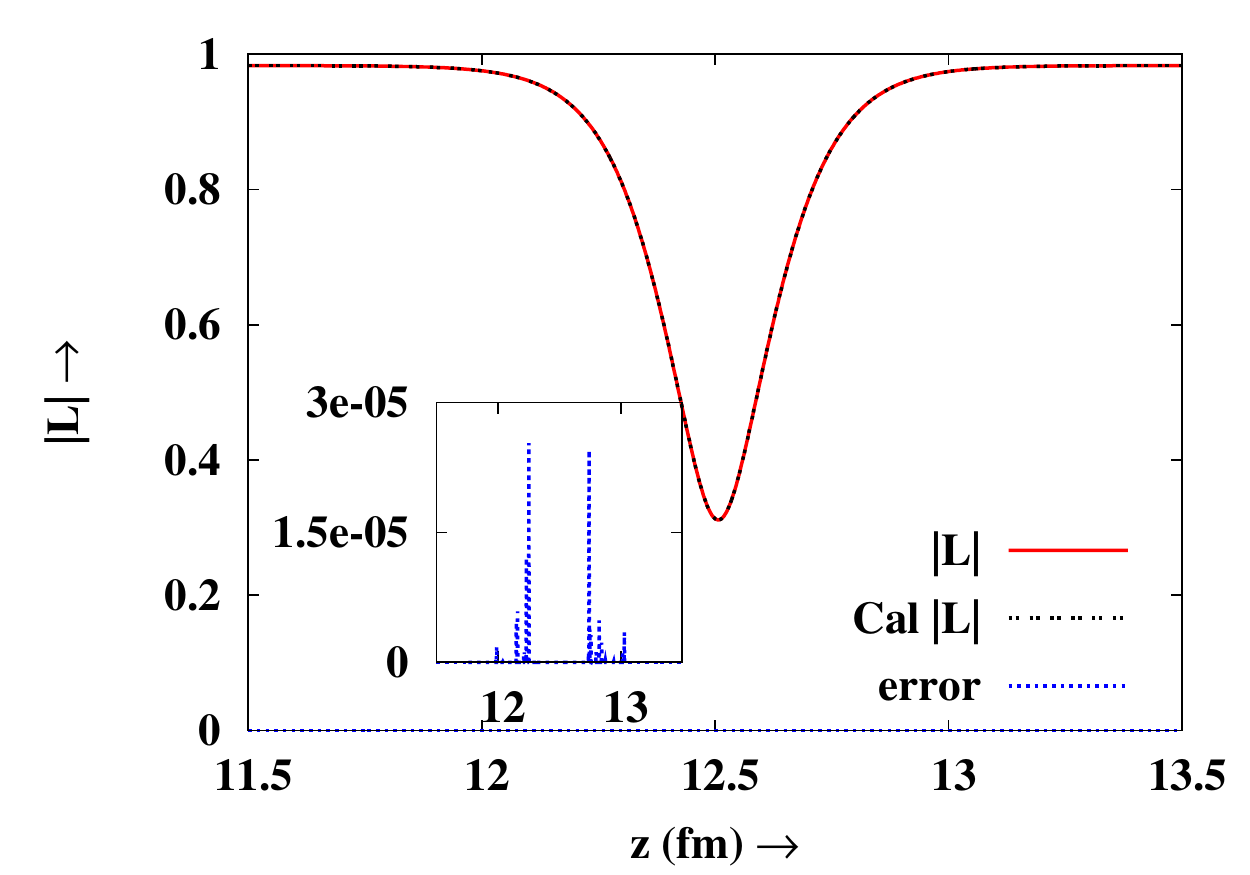}
\includegraphics[width=0.45\textwidth]{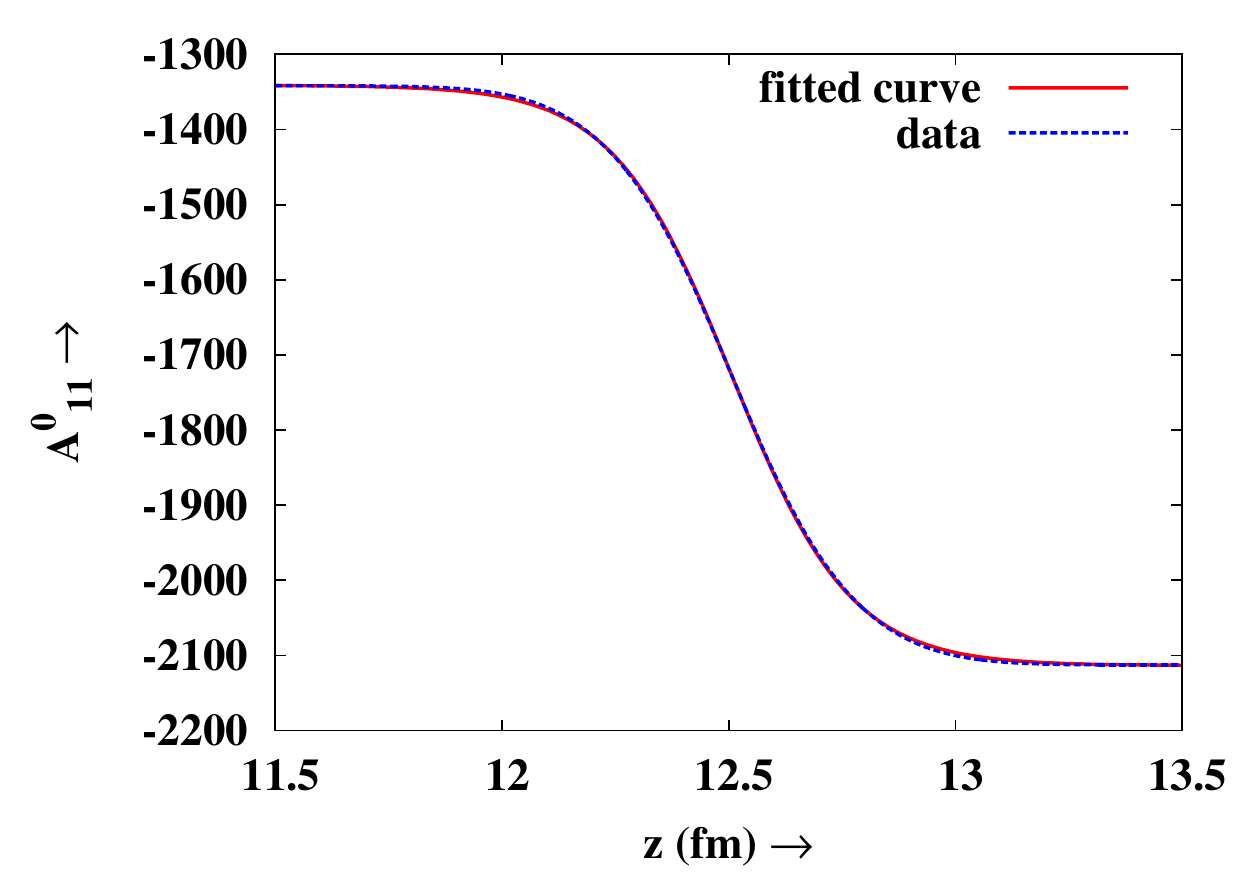}
\caption{Plot of $|l(x)|$ obtained from energy minimization for
$b_1 = 0.0$ and $T=400~MeV$. The corresponding $A_{0}$ configuration is on 
right. $A_{0}$ profile fits well with a tanh function.}
\label{fig:lprflb10}
\end{center}
\end{figure}

  We now include the effects of dynamical quarks leading to explicit
breaking of $Z(3)$ symmetry. For this, we will follow the approach where 
the explicit breaking of the $Z(3)$ symmetry is represented
in the effective potential by inclusion of a linear term in $l$
\cite{Dumitru:2000in,Dumitru:2001bf,Dumitru:2002cf,Dumitru:2003cf}. The above 
potential $V(l)$ with the linear term becomes,

\begin{equation}
 V(l)=\Bigl( -\frac{b_1}{2}(l + l^*) -\frac{b_2}{2} |l|^2- 
{b_3\over 6} (l^3+ {l^{\ast}}^3) +\frac{1}{4}(|l|^2)^2 \Bigr){b_4{T^4}}
\label{eq:vb1l}
\end{equation}

Here coefficient $b_1$ measures the strength of explicit symmetry breaking. 
(In view of Eq.(5), $b_1$ is scaled as $b_1 \rightarrow b_1/y^3$.)
A discussion on various values that $b_{1}$ can have is given in 
\cite{Dumitru:2003cf}. A non-zero value of $b_1$ lifts the degeneracy between 
the three $Z(3)$ vacua. Vacua corresponding to $\theta = 2 \pi/3$ ( $l= z$) 
and $\theta = 4 \pi/3$ ( $l= z^{2}$) remain degenerate, while the true
vacuum with a lower energy corresponds to $l= 1$ ($\theta = 0$). Thus, 
$l = z$ and $l = z^2$ vacua become metastable. The value of $b_1$
can be related to the estimates of explicit $Z(3)$ symmetry breaking 
arising from quark effects which have been discussed in the literature. 
In the high temperature limit, the  estimate of the difference in the potential
energies of the $l = z$ vacuum, and the $l = 1$ vacuum, $\Delta V$, is
given in ref. \cite{Dixit:1991et} as,
\begin{equation}
 \Delta V \sim \frac{2}{3} \pi^{2} T^4 \frac{N_l}{N^3} (N^2 - 2)
\label{eq:delv}
\end{equation}
 where $N_l$ is the number of massless quarks. If we take $N_l = 2$
then $\Delta V \simeq 3 T^4$. For  $T = 400 $ MeV, this value of
$\Delta V$ is obtained if we take the value of $b_1 = 0.645$.
For temperatures of order $T_c$, it is not clear what should be 
the appropriate value of $b_1$. It is entirely possible, that
$b_1$ may be very small near $T_c$. (Possible reasons for taking
very small values of $b_1$ are discussed in detail in ref.
\cite{Gupta:2011ag}.) In view of these uncertainties in the magnitude
of explicit symmetry breaking for temperatures near $T_c$, we will
consider a range of values of $b_1$ including very small value of 
$b_1 = 0.03$, and determine the profile of $l(x)$ and the
associated $A_0$ profile for these values of $b_1$.

\section{PROFILES OF $l(x)$ AND ASSOCIATED GAUGE FIELD CONFIGURATION WITH
EXPLICIT SYMMETRY BREAKING}
 \label{sec:asymprfl}

 The explicit symmetry breaking arising from quark effects will have
important effects on the structure of $Z(3)$ walls. For non-degenerate
vacua, even planar $Z(3)$ interfaces do not remain static, and move
away from the region with the unique true vacuum. Thus, while for the
degenerate vacua case every closed domain wall collapses, for the
non-degenerate case this is not true any more. A closed wall enclosing
the true vacuum may expand if it is large enough so that the surface
energy contribution does not dominate.

 The absence of time independent solutions of the field equations
for $Z(3)$ walls leads to complications in the implementation of
the techniques of ref. \cite{Layek:2005fn} for determination of $l(x)$ profile
for the domain wall which were based on the algorithm of energy 
minimization. In ref.\cite{Layek:2005fn}, correct $l(x)$ profile was obtained
from an initial trial profile by fluctuating the value of $l(x)$ at each
lattice point and determining the acceptable fluctuation which 
lowers the energy (with suitable overshoot criterion etc. as described
in detail in ref.\cite{Layek:2005fn}). For the case without explicit symmetry
breaking, a trial initial configuration of $l(x)$ with appropriate
fixed  boundary conditions (corresponding to the two $Z(3)$ vacua under 
consideration) yielded correct profile of $l(x)$ for the wall within
relatively few iterations. However, with explicit symmetry breaking,
this simple procedure fails as energy can always be lowered by shifting
the wall towards to metastable vacua (thus expanding the region with
true vacuum). 

\begin{figure}[!htp]
\begin{center}
\includegraphics[width=0.45\textwidth]{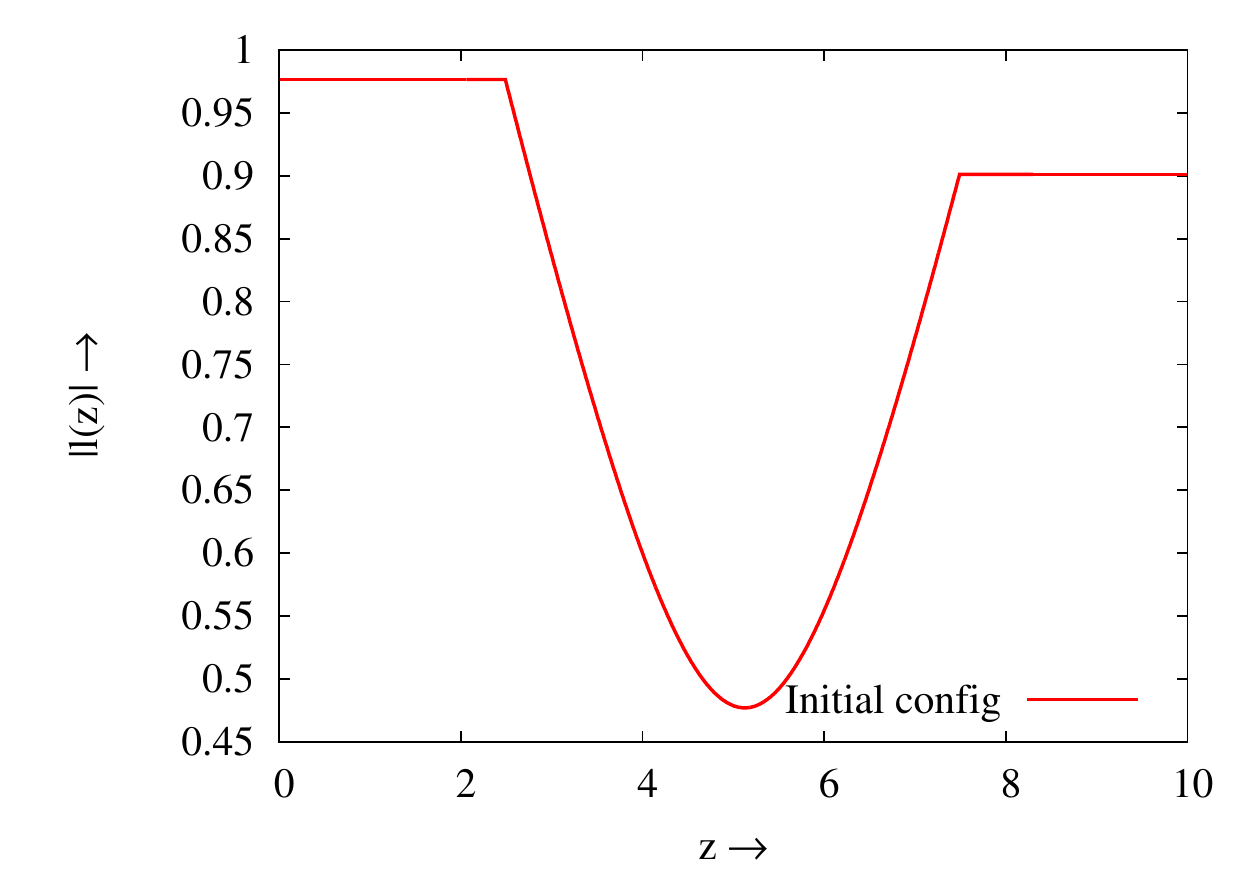}
\includegraphics[width=0.45\textwidth]{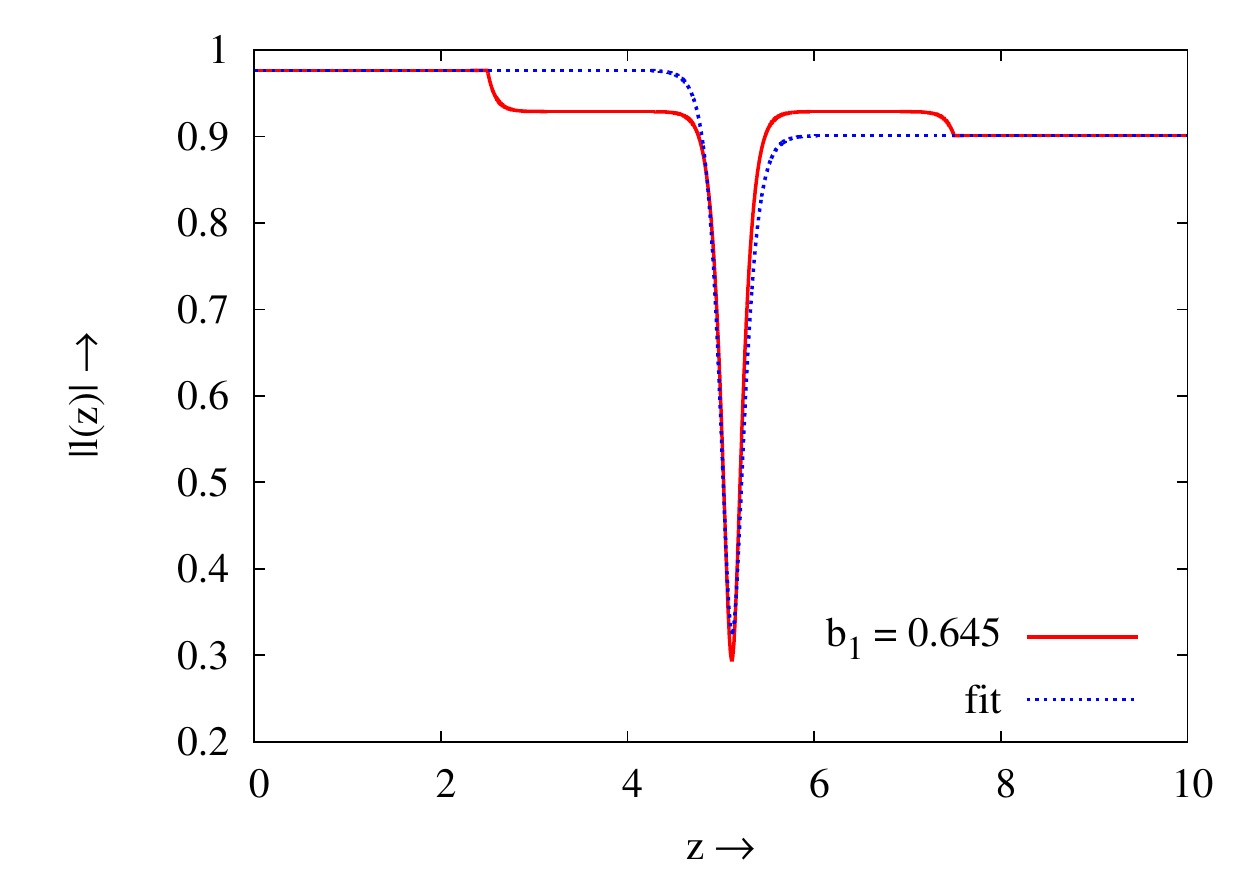}
\caption{Plot of $|l(x)|$ obtained from energy minimization for
$b_1 = 0.645$ (solid curve). On the left is the initial trial 
configuration. The final configuration is on right.}
\label{fig:lprfl}
\end{center}
\end{figure}

From the computational point of view, one of the 
major change due to the inclusion of $b_{1}$ term is the the scaling. Without 
$b_{1}$ all the vacua are degenerate, so $|l(x)| \rightarrow 1$ in all the 
vacua. However, that is not the case with the potential given by Eq. 
(\ref{eq:vb1l}). This leads to the $b_{1}$ dependence of the scaling. We 
normalize the potential in such a manner that $|l(x)| \rightarrow 1$ in the 
true vacuum. As we mentioned above, the energy splitting between vacua 
itself amounts to a pressure difference between the two vacua.  Thus the 
program tries to minimize the energy by moving the domain wall in one 
direction till it goes completely out of the lattice, in the process it 
changes the boundary values too if they are not held fixed. If we fix the 
boundary value in the far left and far right region of lattice, the program 
minimizes the energy by not only moving the profile in the intermediate 
region but also by re-adjusting the values of $|l(x)|$ on 
the two sides. The effect is most pronounced 
for the large $b_{1}$ .This statement becomes clearer if we look at the Fig. 
\ref{fig:lprfl}. It shows the initial and the final profile of $l(x)$ 
between $l=1$ and $l=z$ vacua for $b_1 = 0.645$ at $T = 400~MeV$. The 
asymmetry is pretty clear in the boundary conditions of the initial trial 
configuration itself. Note the 
central region in the final configuration (solid curve). There is a sharp 
variation of $|l(x)|$ in a small region and on either side of it the $|l(x)|$ 
values are same (but different from actual boundary values) leading to a 
stable configuration in the middle. Since the domain wall is characterized by 
the sharp variation of the field in a small spatial region, we fit the profile 
such that it meets the correct boundary values while keeping the variation 
as given by the energy minimization program. This is shown by the dotted curve 
in the left figure. Though this procedure of {\it smoothening} the
domain wall profile near its edges is somewhat ad hoc, it will not affect 
our results much as the scattering of quarks and antiquarks are primarily
decided by the height and width of the sharply varying profile of $l(x)$.
 On comparing with Fig. (\ref{fig:lprflb10}) 
(for $b_1 = 0$ case), we note that explicit breaking of $Z(3)$ symmetry 
leads to asymmetric profiles of $l(x)$. This immediately suggests that 
there will be a difference between the scattering  of a quark coming from 
the right and the scattering of the one coming from left.

\begin{figure}[!htp]
\begin{center}
\includegraphics[width=0.45\textwidth]{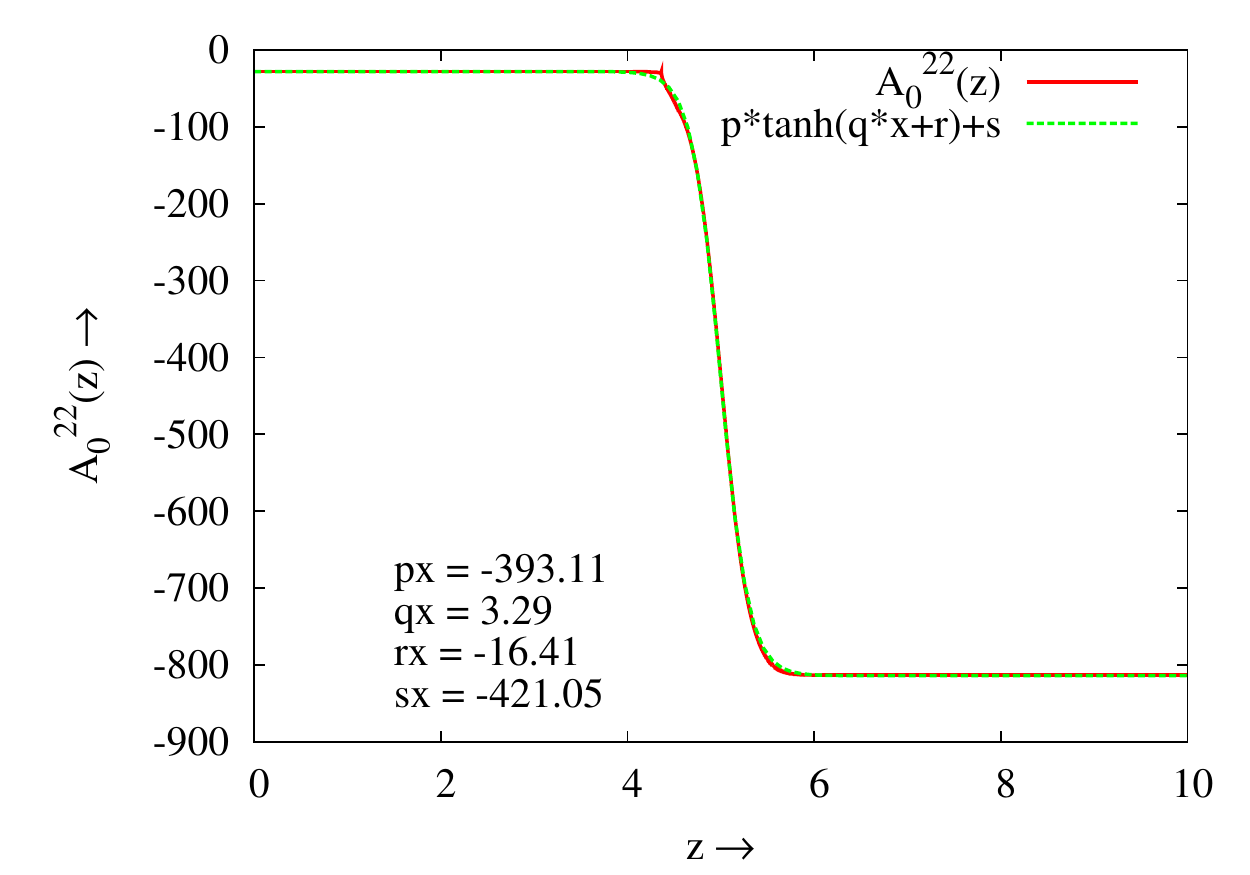}
\includegraphics[width=0.45\textwidth]{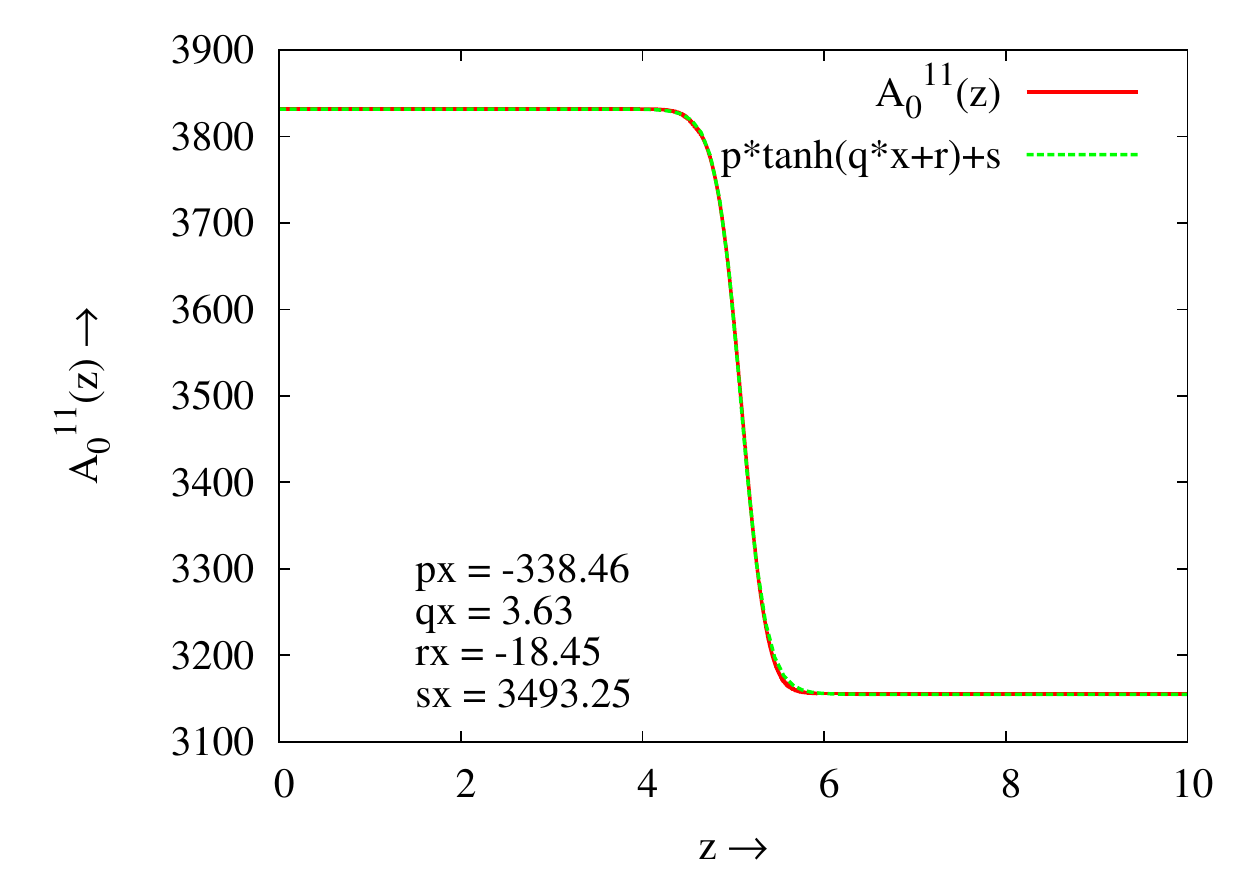}
\caption{Plot of calculated $A_{0}$ and the fitted profile ($A_{0}(x) = 
p \tanh (qx + r) + s$) for $b_{1} = 0.03$ and $0.645$.}
\label{fig:a0plt}
\end{center}
\end{figure}

  The $A_{0}$ profile corresponding to the $l(x)$ profile was
calculated in our earlier paper \cite{Atreya:2011wn}, where we also discussed 
various conceptual issues related to the ambiguities in the extraction 
of a colored quantity $A_0$ from color singlet $l(x)$. We choose Polyakov 
gauge (diagonal gauge) for $A_{0}$: 
\begin{equation} 
A_{0} = \frac{2\pi T}{g}\left(a\lambda_{3} + b\lambda_{8}\right),
\label{eq:a0diag}
\end{equation}
where, $g$ is the coupling constant and $T$ is the temperature, while
$\lambda_{3}$ and $\lambda_{8}$ are the diagonal Gell-Mann matrices.
The $A_{0}$ profile was obtained from $l(x)$ profile
(Fig. \ref{fig:lprfl}) by inverting Eq.(\ref{eq:lx}).
For details, see ref. \cite{Atreya:2011wn}. We have carried out this
calculation for the profiles of $l(x)$ obtained from the energy
minimization program for $b_{1}\neq 0$(Fig. (\ref{fig:lprfl})).
The calculated $a,b$ were then used to calculate $A_{0}$ using  
Eq. (\ref{eq:a0diag}). The $A_{0}$ profile thus obtained is reasonably well 
fitted to the function $A_{0}(x) = p \tanh (qx + r) + s$ using 
gnuplot. The calculated $A_{0}$ profile and fitted
$A_{0}$ profile are plotted in figure (\ref{fig:a0plt}).

 We note that the fit to $\tanh$ profile is almost perfect just as
was the case for $b_1 = 0$ case in Fig. \ref{fig:lprflb10}. We thus conclude 
that the scattering of a quark coming from left with such an $A_0$ profile 
(in the Dirac equation) will be the same as the scattering of an antiquark 
coming from right (with same kinetic energy). Thus a collapsing domain wall 
with $l = 1$ inside and $l = z$ outside will give same reflection coefficients 
(hence resulting concentration) for quarks inside as a collapsing domain
wall with $l = z$ inside and $l = 1$ outside will give for antiquarks
(assuming zero baryon chemical potential). This is 
interesting in view of the asymmetric profiles of $l(x)$ in 
Fig. \ref{fig:lprfl}  for $b_1 \neq 0$ cases. Though still there will
be important differences from the $b_1 =0$ case as now a sufficiently
large closed domain wall with true vacuum ($l = 1$) inside will expand
instead of collapsing, leading to concentration of quarks or antiquarks
in a shell like region. We will discuss these possibilities later in
section \ref{sec:discussion}. 

 It may also be noted that we have shown $A_{0}^{11}$ for $b_{1}=0.645$ and 
$A_{0}^{22}$ for $b_{1}=0.03$. This is for the reason that both the profiles 
are similar in the shape and size. It has to do with the choice of initial 
$(a,b)$ values while calculating $A_{0}$. This essentially means that we 
should compare the reflection of red quark in $b_{1} = 0.645$ case with the 
reflection of green quark in $b_{1}=0.03$ case. One may use hit and trial 
method to find a specific choice of $(a,b)$ in the case of $b_{1} = 0.03$ such 
that $A_{0}^{11}$ obtained has the same spatial variation as the one for 
$b_{1}=0.645$. We refer to ref. \cite{Atreya:2011wn} for further details on 
this issue of initial conditions.

   As we mentioned, it is interesting to note that asymmetry of $l(x)$ is not 
reflected in the background gauge configuration. 
The effect of non-zero $b_{1}$ is reflected in the $A_{0}$ profile not in 
terms of the change in shape but in terms of the height of the potential 
getting reduced. For $b_{1}= 0.645$, the height of $A_{0}$ is almost $100~MeV$ 
less than the height of $A_{0}$ in $b_{1} = 0.03$ case. However, this 
decrease in the height will not give any asymmetry in the reflection of quarks 
and anti-quarks from the $A_{0}$, neither will it change the amount of 
reflection in a drastic fashion. We will now consider another possibility 
which allows for asymmetry in concentration of quarks and antiquarks for the 
$b_1 \neq 0$ case.

  For this we recall the discussion of quark/antiquark scattering due
to $l$ dependent effective mass, as discussed in ref.\cite{Layek:2005zu}.
The basic idea proposed in ref.\cite{Layek:2005zu} was that as $l(x)$ is the 
order parameter for the quark-hadron transition, physical properties such 
as effective mass of the quarks should be determined in terms of $l(x)$.  
This also looks natural from the expected correlation between the chiral 
condensate and the Polyakov loop. Lattice results indicate that the chiral 
phase transition and the deconfinement phase transition may be coupled,
i.e as the Polyakov loop becomes non zero across $T_{c}$, the chiral order 
parameter attains a vanishingly small value. Thus, if there is spatial 
variation in the value of $l(x)$ in the QGP phase then effective mass of 
the quark traversing that region should also vary (say, due to spatially 
varying chiral condensate). For regions
where $l(x) = 0$, quarks should acquire constituent mass as appropriate
for the confining phase. To model the dependence of effective quark mass on 
$l(x)$ we could use the color dielectric model of ref.\cite{Phatak:1998iv} 
identifying $l(x)$ with the color dielectric field $\chi$ in 
ref.\cite{Phatak:1998iv}. Effective mass of the quark was modeled in 
\cite{Phatak:1998iv} to
be inversely proportional to $\chi$. This leads to divergent quark mass
in the confining phase consistent with the notion of confinement.
However, we know that the divergence of quark energy in the confining
phase should be a volume divergence (effectively the length of string
connecting the quark to the boundary of the volume). $1/l(x)$ dependence
will not have this feature, hence we do not follow this choice. For the
sake of simplicity, and for order of magnitude estimates at this
stage, we will model the quark mass dependence on $l(x)$ in the following
manner.

\begin{equation}
m(x) = m_q + m_0(l_0 - |l(x)|)
\label{eq:effm}
\end{equation}

Here $l(x)$ represents the profile of the $Z(3)$ domain wall, and $l_0$ is 
the vacuum value of $|l(x)|$ (for the true vacuum) appropriate for the 
temperature under consideration. $m_q$ is
the current quark mass of the quark as appropriate for the QGP
phase with $|l(x)| = l_0$, with $m_u \simeq m_d = 10$ MeV and $m_s 
\simeq 140$ MeV. $m_0$ characterizes the constituent
mass contribution for the quark. We will take $m_0 = 300$ MeV. Note
that here $m(x)$ remains finite even in the confining phase with
$l(x) = 0$. As mentioned above, this is reasonable since we are
dealing with a situation where $l(x)$ differs from $l_0$ only in
a region of thickness of order $1 fm$ (thickness of domain wall). 

The space dependent part of $m(x)$ in Eq.(\ref{eq:effm}) is taken as a 
potential term in the Dirac equation for the propagation of quarks and 
antiquarks. As we see from Fig.(\ref{fig:lprflb10}), $l(x)$ varies across
a $Z(3)$ interface, acquiring small magnitude in the center of the wall.
A quark passing through this interface, therefore, experiences a nonzero
potential barrier leading to non-zero reflection coefficient for
the quark. Important thing here is that due to asymmetric profile of
$l$ (Fig.(\ref{fig:lprfl})), the effective mass of quarks/antiquarks will 
have different
values on the two sides of the domain wall. This effect, when combined with
the scattering from the background $A_0$ configuration, will lead to asymmetry
in the scattering of quarks from one side and that of antiquarks from
the other side of the domain wall. 

One may be concerned here whether 
combining the scattering from $A_0$ configuration with the scattering due
to $l$ dependent effective mass amounts to double counting in the sense that
both effects originate from the same $l(x)$ profile. For this we note that
there are indeed two different effects at play here due to the existence
of $Z(3)$ walls. First effect arises from the existence of three different 
phases of QGP characterized by spontaneous breaking of $Z(3)$ symmetry.
In the absence of explicit symmetry breaking one will expect that physics
should be identical for these three phases. Thus, even $l$ dependent effective
mass of quarks should have the same value in these three phases, as indeed
is the case from Eq.(9) due to same value of $|l|$ in the three 
$Z(3)$ phases. However, with explicit symmetry breaking, there is no physical
argument to say that physics should be the same for the three $Z(3)$ vacua,
as the two vacua ($l = z$ and $l = z^2$) become metastable. As $|l|$ in
these two vacua has smaller magnitude, effective mass of quarks may actually
be larger in these two phases of QGP. As explained for Eq.(9), we can think
of this $|l|$ dependent mass in terms of chiral condensate whose value
will depend on $l(x)$.  (We mention that $l(x)$ dependent quark mass by 
itself is a non-trivial implication of our proposal and it will have many other 
interesting implications on propagation of quarks/antiquarks in QGP in the 
presence of these $Z(3)$ domains.) Next we come to the presence of background 
gauge field. This arises from 
spatial variation of $l(x)$ leading to color electric field from which quarks 
and antiquarks scatter in different manner. This color electric field is 
entirely localized at the boundary of $Z(3)$ domains (where $l(x)$ has 
spatial variation), and vanishes in the interiors of the $Z(3)$ domains. It
couples differently to quarks/antiquarks of different color charges. Hence, 
this effect is entirely different from the effect of effective mass which has 
different values in the  interiors of the two domains, irrespective of the
color charges of quarks and antiquarks (even though for the 
scattering purposes, both effects lead to non-trivial potential at the 
location of the $Z(3)$ wall).   

\section{Reflection and Transmission Coefficients with explicit 
symmetry breaking}
\label{sec:ref-trans}

We now calculate the reflection and transmission coefficient for quarks
and antiquarks subject to the above two effects. One is CP violating, arising 
from 
the background gauge field $A_0$ (Eq.(\ref{eq:a0diag})), and the other is CP
preserving, arising from the space dependent effective mass of 
quarks/antiquarks (Eq.(\ref{eq:effm})). We recall the steps for calculation 
from \cite{Atreya:2011wn}.
To calculate the reflection and transmission coefficient, we need the
solutions of Dirac equation in the Minkowski space but the $A_{0}$ profile is
calculated in Euclidean space. We start with the Dirac equation in the
Euclidean space, with the spatial dependence of $A_0$ calculated from Z(3)
wall profile as mentioned above, and with space dependent mass term as given
in Eq.(9). 
\begin{equation}
\label{eq:eucliddiraceq}
\bigl[\gamma^{0}_{e}\partial_{0}\delta^{jk}  - g\gamma^{0}_{e}A_{0}^{jk}(z) 
+ (i \gamma^{3}_{e}\partial_{3} + m(x))\delta^{jk}\bigr]\psi_{k} = 0,
\end{equation}
where $\gamma^{0}_{e} \equiv i\gamma^{0}$ and $\gamma^{3}_{e} \equiv 
\gamma^{3}$ are the Euclidean Dirac matrices. $\partial_0$ denotes 
$\partial/\partial_\tau$ with $\tau = it$ being the Euclidean time.
$j,k$ denote color indices. $m(x)$ is the effective mas as given in 
Eq.(9). We now analytically continue the Eq. (\ref{eq:eucliddiraceq}) to 
the Minkowski space to get
\begin{equation}
\label{eq:minkdiraceq}
\bigl[i\gamma^{0}\partial_{0}\delta^{jk} + g\gamma^{0}A_{0}^{jk}(z) + 
(i\gamma^{3}\partial_{3} + m(x)\bigr)\delta^{jk}]\psi_{k} = 0.
\end{equation}
where now $\partial_0$ denotes $\partial/\partial t$ in the Minkowski
space.

 Eq.(\ref{eq:minkdiraceq}) is used to calculate the reflection and 
transmission coefficients. For a general smooth potential we
followed a numerical approach given by Kalotas and Lee \cite{Kalotas:1991kl}.
They have discussed a numerical technique to solve Schr\"{o}dinger equation
with potentials having arbitrary smooth space dependence. We applied this
technique of ref.\cite{Kalotas:1991kl} for solving the Dirac equation (see,
ref.\cite{Atreya:2011wn} for details).

 The results for different quarks and anti-quarks (with $E$ = 3.0 GeV 
taken as example for each case) are given in table \ref{tab:refquarks}. As we 
mentioned, the important quantity for us is to calculate the reflection 
coefficient of (say) quarks coming from the left of the wall and compare it 
with the reflection coefficient of antiquarks (with the same kinetic energy) 
coming from the right of the wall. Any (possible) difference in these two 
reflection coefficients directly relates to the expected concentration
of quarks and antiquarks by a domain wall of one kind and its opposite 
wall (interpolating between the two $Z(3)$ vacua in reverse order).
Table \ref{tab:refquarks} shows clear difference in these two reflection 
coefficients.
\begin{table}[!htp]
\begin{center}
\begin{tabular}{|c|c|c|c|}
\hline
      & $b_{1}=0.03$ & $0.126$ & $0.645$ \\
     \hline
      Left $R_{q}$ & $1.65437\times10^{-6}$ & $4.40706\times10^{-6}$ & $1.43314
\times10^{-10}$\\
      \hline
     Right $R_{q}$ & $0.00003366$ & $0.0141752$ & $0.00394808$\\
     \hline
      Left $R_{aq}$ & $2.25671\times10^{-6}$ & $1.85367\times10^{-7}$ & $2.07835
\times10^{-7}$\\
     \hline
      Right $R_{aq}$ & $0.000376883$ & $0.0820803$ & $0.073885$\\
    \hline
\end{tabular}
\caption{Table for the reflection coefficients for various quarks and 
antiquarks for smooth profiles of $A_0$ and $m(x)$.}
\label{tab:refquarks}
\end{center}
\end{table} 

\section{Discussion}
\label{sec:discussion}

 In this work we have extended our earlier studies of CP violating 
scattering of quarks/antiquarks from Z(3) walls 
\cite{Atreya:2011wn,Atreya:2014sea,Atreya:2014sca},
by including the effects of explicit breaking of Z(3) symmetry which is 
expected to arise due to dynamical quarks. The resulting profile of $l(x)$ 
between the true vacuum and a metastable vacuum is no more symmetric 
in this case which leads to new effects. We study scattering of quarks
and antiquarks from the background $A_0$ field associated with the
profile of $l(x)$ while also incorporating the effect of spatially varying
effective mass of quarks and antiquarks in the respective $Z(3)$ domains.
The combined effect of the scattering shows interesting behavior leading 
to left-right asymmetry in scattering of quarks (from left) and antiquarks 
(from right). This will lead to important differences in resulting 
concentrations of quarks and antiquarks in cosmology as well as in RHICE.
For example, in the early universe, a network of domain walls will arise
with varying sizes and interpolating between different $Z(3)$ vacua. For
all domain walls of a given size interpolating between given two vacua
in a given order, there will be roughly same number of walls with 
similar size but interpolating between the same two $Z(3)$ vacua in the
reverse order. (Though explicit symmetry breaking may also produce difference
between formation of such walls, introducing further richness in the effects
of explicit symmetry breaking). In the absence of explicit symmetry breaking, 
if first type of walls give certain concentration of (say) quarks, then
the other set of walls will give similar concentration of antiquarks.
This is, however, not the case when explicit symmetry breaking effects
are incorporated. In view of results from table \ref{tab:refquarks}, the two 
sets of
walls will lead to very different concentrations of quarks and antiquarks
(especially if the value of $b_1$ is large). Though for each domain wall
(say interpolating between $l = 1$ and $l = z$, there is always the
{\it conjugate} wall (interpolating between $l = 1$ and $l = z^2$)
which will lead to same scattering between quarks and antiquarks.
Final effect of our results will then appear as two different 
magnitudes for the concentrations of quarks and antiquarks, even if one
takes all domain walls of the same size. This is very different from the
case without explicit symmetry breaking where domain walls of same size
will lead to quark and antiquark inhomogeneities of same magnitude (for
same kinetic energies of quarks and antiquarks). This difference will
be particularly dramatic for RHICE where number of domain walls if
of order one for each event \cite{Gupta:2010pp}. Thus even for same type of
events, one may get very different concentration of baryons or antibaryons
in different events leading to very large event-by-event fluctuations.  

Situation is even more interesting when we consider the effect that
with explicit symmetry breaking certain closed domain walls may expand, 
those with true vacuum inside (and with sufficiently larger size
so that volume energy difference dominates over the surface energy
contribution \cite{Gupta:2011ag}). This can lead to concentration of 
quarks and antiquarks in a shell like  structure. For cosmology very large
expanding domain walls may trap shells of baryons/antibaryons if enclosed
by a collapsing {\it antiwall} configuration. Such shells can form in RHICE
also and will have important observations signatures.
 
\section{Acknowledgments}
 We are extremely thankful to Shreyash Shankar Dave, Sanatan Digal, and 
Saumia P.S. for fruitful discussions.

%\bibliographystyle{apsrev}
%\bibliography{z3cpqrk}

\end{document}